\documentclass[a4paper]{article}
\usepackage{amsmath}
\usepackage{amsfonts}
\usepackage[T1]{fontenc}
\usepackage[english]{babel}
\usepackage[latin1]{inputenc}
\usepackage{url}
\usepackage{ae}
\usepackage{aecompl}
\usepackage{amsthm,amssymb,upref,amscd}
\newtheorem{theorem}{Theorem}
\newtheorem{lemma}[theorem]{Lemma}

\title{Calculation of, and bounds for, the multipole moments of stationary spacetimes}
\author{Thomas B\"ackdahl${}^*$, Magnus Herberthson\thanks{Department of  
Mathematics, Link\"oping University,
SE-581 83 Link\"oping, Sweden.\newline
\hspace*{5mm} e-mail: thbac@mai.liu.se, maher@mai.liu.se}}
\date{}

\begin{document}
\maketitle
\begin{abstract}
In this paper the multipole moments of stationary asymptotically flat
spacetimes are considered. We show how the tensorial recursion of Geroch and Hansen can
be replaced by a scalar recursion on $\mathbb{R}^2$.
We also give a bound on the multipole moments.
This gives a proof of the "necessary part" of a long standing conjecture due to Geroch.
\end{abstract}
\section{Introduction}
The relativistic multipole moments of asymptotically flat spacetimes have been defined by Geroch \cite{geroch} for static spacetimes and then generalised to 
the stationary case by Hansen \cite{hansen}. Together with Beig's \cite{beigAPA} generalised definition of centre of mass this gives
a coordinate independent description of all asymptotically flat stationary spacetimes. 

However, the tensorial recursion which defines the multipole moments \eqref{orgrec}, is computationally rather complicated as it stands. In the axisymmetric (static or stationary) case, on the other hand, it was shown,
\cite{backdahl1}, \cite{backdahl2} that the recursion can be replaced
by a scalar recursion on $\mathbb{R}$, and that all the moments can be collected
into one complex valued function $y$ on $\mathbb{R}$, where the moments are given by the
the derivatives if $y$ at $0$.

In the general case, the multipole of order $2^n$ has $2n+1$ degrees of freedom, as compared to one degree of freedom in the axisymmetric case. Therefore, apart from the technical problems, it is not obvious what form a generalisation to the general case should take. In this paper, we show that also in the general stationary case, the recursion
\eqref{orgrec} can be simplified to a scalar recursion, this time on
$\mathbb{R}^2$. This is shown using normal coordinates, complex null geodesics, and exploiting the extra conformal freedom of the conformal compactification. 

Using this simplification we can partially confirm an extension of a long standing conjecture by Geroch \cite{geroch}:
\begin{quote}
Given any set of multipole moments, subject to the appropriate convergence condition, 
there exists a static solution of Einstein's equations having precisely those moments.
\end{quote}
This conjecture has its natural extension to the stationary case.

In this paper we will state the appropriate convergence condition in the general stationary case,
i.e., we will prove that this condition is necessary for existence of a stationary solution to Einstein's equations. 

\section{Multipole moments of stationary spacetimes} \label{tmoments}
In this section we quote the definition of multipole moments given by Hansen in \cite{hansen}, which is an extension 
to stationary spacetimes of the definition by Geroch \cite{geroch}.
We thus consider a stationary spacetime $(M, g_{ab})$ with timelike Killing vector field $\xi^a$.
We let $\lambda=-\xi^a\xi_a$ be the norm, and define the twist $\omega$ through 
$\nabla_a\omega=\epsilon_{abcd}\xi^b\nabla^c\xi^d$.
If $V$ is the 3-manifold of trajectories, the metric $g_{ab}$ (with signature $(-,+,+,+)$) induces the 
positive definite metric
$$h_{ab}=\lambda g_{ab}+\xi_a\xi_b$$ on $V$.
It is required that $V$ is asymptotically flat, i.e.,  there exists a 3-manifold
$\hat V$ and a conformal factor $\Omega$ satisfying
\begin{itemize}
\item[(i)]{$\hat V = V \cup \Lambda$, \; where $\Lambda$ is a single point}
\item[(ii)]{$\hat h_{ab}=\Omega^2 h_{ab}$ is a smooth metric on $\hat V$}
\item[(iii)]{At $\Lambda$, $\Omega=0, \hat D_a \Omega =0, \hat D_a \hat D_b \Omega = 2 \hat h_{ab}$,}
\end{itemize}
where $\hat D_a$ is the derivative operator associated with $\hat h_{ab}$.
On $M$, and/or $V$ one defines the scalar potential
$$\phi=\phi_M+i\phi_J, \quad \phi_M=\frac{\lambda^2+\omega^2-1}{4\lambda}, \,\phi_J=\frac{\omega}{2\lambda}.$$
The multipole moments of $M$ are then defined on $\hat V$ as certain  
derivatives of the scalar
potential $\hat \phi=\phi/\sqrt \Omega$ at $\Lambda$. More  
explicitly, following \cite{hansen}, let $\hat R_{ab}$ denote
the Ricci tensor of $\hat V$, and let $P=\hat \phi$. Define the  
sequence $P, P_{a_1}, P_{a_1a_2}, \ldots$
of tensors recursively:
\begin{equation} \label{orgrec} P_{a_1 \ldots a_n}=C[
\hat D_{a_1}P_{a_2 \ldots a_n}-
\tfrac{(n-1)(2n-3)}{2}\hat R_{a_1 a_2}P_{a_3 \ldots a_n}],
\end{equation}
where $C[\ \cdot \ ]$ stands for taking the totally symmetric and  
trace-free part. The multipole moments
of $M$ are then defined as the tensors $P_{a_1 \ldots a_n}$ at  
$\Lambda$. The requirement that all $P_{a_1 \ldots a_n}$ be totally symmetric and
trace-free makes the actual calculations very cumbersome.

In \cite{beigPRSL}, \cite{kundu} it was shown that (when the mass is non-zero) there exist a conformal 
factor $\Omega$ and a chart, such that all components of the metric $\hat h_{ab}$ and the potential $\hat \phi$ are analytic in terms of the 
coordinates, in a neighbourhood of the infinity-point. Expressed in these coordinates, the exponential map becomes analytic. 
Therefore, we can use Riemannian normal coordinates and still have analyticity of the metric components and the potential.
If the mass is zero, this analyticity condition will be assumed. Thus, henceforth we assume that $\Omega$ is chosen such that
the (rescaled) metric and potential are analytic in a neighbourhood of $\Lambda$.

\section{Multipole moments through a scalar recursion on $\mathbb{R}^2$}
\label{momentsscalar}
Suppose that $(x^1, x^2, x^3)=(x, y, z)$ are normal coordinates (with respect to 
$\hat h_{ab}$) centred around $\Lambda$.
This means that for any constants $a=a^1, b=a^2 , c=a^3$ the curve $t \to (a t,b t,c t)$ is a geodesic,
i.e. in terms of coordinates that
$$ \ddot x^i+\Gamma^i_{kl}\dot x^k \dot x^l=\Gamma^i_{kl}a^k a^l=0$$
where the Christoffel symbols are evaluated at $(a t,b t,c t)$ for appropriate $t$. Due to analyticity,
this relation holds for complex values of $a,b,c$, i.e., we can consider geodesics in the complexification
$\hat V_\mathbb{C}$ of $\hat V$. Of particular interest is the one-parameter family of curves:
$$\gamma_\varphi: t \to (t \cos\varphi,t \sin \varphi, i t),
 \qquad t \in [0,t_0), \varphi\in [0,2\pi)$$
 for some suitable $t_0$. The tangent vector $\eta^a=\eta^a_\varphi(t)=
\cos\varphi (\frac{\partial}{\partial x^1})^a+\sin\varphi (\frac{\partial}{\partial x^2})^a+i (\frac{\partial}{\partial x^3})^a$ is seen to be a complex null vector along $\gamma_\varphi$.
Namely, from $\eta^a \hat D_a \eta^b=0$, we infer that $\eta^a \hat D_a (\eta^b \eta_b)=0$.
The (constant) value of $\eta^b \eta_b$ is then found to be $0$ by evaluation at $t=0$.
 
Next, consider the mapping $F: \mathbb{R}^2 \to V_\mathbb{C}: 
(\xi,\zeta) \to (\xi,\zeta,i \sqrt{\xi^2+\zeta^2})$. We let $S$ denote the 2-surface $F(U) \subset \hat V_\mathbb{C}$, where $U \subset \mathbb{R}^2$ is a suitable neighbourhood of $(\xi,\zeta)=(0,0)$. 
$S$ is then a smooth surface, except at $\Lambda$ where it has a vertex point, closely resembling
a null cone in a three dimensional Lorentzian space.
The curves $\gamma_\varphi$ are given by $\gamma_\varphi(t)=F(t\cos\varphi,t\sin\varphi)$, and
in particular $\eta^a$ lies along $S$.
This suggests that we use the polar coordinates $\rho,\varphi$ around $\Lambda$ on $S$ defined via 
$\xi=\rho\cos\varphi$, $\zeta=\rho\sin\varphi$.
We now follow the approach from \cite{backdahl2}, where a useful vector field $\eta^a$ on $\hat V$ was introduced. 
In \cite{backdahl2}, where the spacetime was axisymmetric, $\eta^a$ was explicitly expressed in 
terms of the metric cast in the Weyl-Papapetrou form \cite{wald}, and was defined on the whole of $\hat V$ except on the symmetry axis. 
In this paper the axisymmetry condition is dropped, which makes the construction of a corresponding $\eta^a$ more difficult. 
In addition, a general spacetime has $2n+1$ degrees of freedom for the multipole moment of order $2^n$, 
compared to one degree of freedom in the axisymmetric case \cite{herb}.
Nevertheless, it will turn out to be sufficient to know the potential $\hat \phi$ on $S$ to determine all the moments. 
On $S$, $\eta^a$ has the following properties.

\begin{lemma}\label{etalemma}
Suppose $\hat V$ and $S$ are defined as above. Then there exists a
regularly direction dependent (at $\Lambda$) vector field $\eta^a$ on $S$ with the following properties:\\
a) $\eta^a  \hat D_a \eta^b$ is parallel to $\eta^b$.\\
b) For all tensors $T_{a_1 \ldots a_n}$, $\eta^{a_1} \ldots
\eta^{a_n}T_{a_1 \ldots a_n}=\eta^{a_1} \ldots \eta^{a_n}C[T_{a_1
  \ldots a_n}]$,\\
c) At $\Lambda$, $P_{a_1 \ldots a_n}$ (in $\hat V$) is determined by 
$\eta^{a_1} \ldots \eta^{a_n}P_{a_1 \ldots a_n}$ (on $S$)
\end{lemma}

\begin{proof}
a) was demonstrated above, b) follows as in \cite{backdahl2}, while c) requires a different argument.
A totally symmetric and trace-free tensor $P_{a_1\dots a_n}$ has $2n+1$ degrees of freedom, and
in Cartesian coordinates $(x,y,z)$ it can be expressed via the components 
$P\underbrace{{}_{x  \ldots\ldots x}}_{j}\underbrace{{}_{y ..\ldots\ldots y}}_{n-j-1}{}_z$
and $P\underbrace{{}_{x\ldots\ldots x}}_{j}\underbrace{{}_{y\ldots\ldots y}}_{n-j}$.
\footnote{In brief, any index occurrence of several z's can be removed via $P_{zz\dots}+P_{xx\dots}+P_{yy\dots}=0$. }
Therefore, at $\Lambda$, we can write
\begin{equation}
\label{reconstrtensor}
\begin{split}
P_{a_1\dots a_n}=
\sum_{j=0}^n{a_{j}C[(dx)_{a_1}\dots(dx)_{a_j}(dy)_{a_{j+1}}\dots(dy)_{a_{n}}]}\\
+\sum_{j=0}^{n-1}{b_{j}C[(dx)_{a_1}\dots(dx)_{a_j}(dy)_{a_{j+1}}\dots(dy)_{a_{n-1}}(dz)_{a_{n}}]}
\end{split}
\end{equation}
Contracting with $\eta^{a_1}\dots\eta^{a_n}$, and using lemma \ref{etalemma}b we find that
\begin{equation}\label{trigpol}
\eta^{a_1}\dots\eta^{a_n}P_{a_1\dots a_n}=\sum_{j=0}^n{a_{j}\cos^j\varphi\sin^{n-j}\varphi}
+i\sum_{j=0}^{n-1}{b_{j}\cos^j\varphi\sin^{n-1-j}\varphi}
\end{equation}
If the left hand side is zero, the trigonometric polynomial to the right must be identically zero.
This means that all the coefficients $a_n$ and $b_n$ are zero, and by \eqref{reconstrtensor}
that $P_{a_1\dots a_n}$ is zero. In particular the $2n+1$ components in the RHS of
\eqref{reconstrtensor} are linearly independent. This proves c).
\end{proof}
Note that although the moments are encoded in the coefficients $a_n$ and $b_n$, this encoding
is dependent on the choice of normal coordinates, i.e., the orientation of the coordinate axes in $T_\Lambda \hat V_\mathbb C$.

We can now replace the recursion \eqref{orgrec} on  $\hat V$ with a scalar recursion on $S$.
Again, we follow \cite{backdahl2}, and define
\begin{equation} \label{fn}
f_n=\eta^{a_1}\eta^{a_2}\dots\eta^{a_n}P_{a_1a_2\dots a_n}, \ n=0,1,2,\ \ldots
\end{equation}
on $S$. In particular, $f_0=P=\hat\phi=\phi/\sqrt{\Omega}$. The moments 
$P_{a_1a_2\dots a_n}(\Lambda)$ will now be encoded in the trigonometric polynomial
given by the direction dependent limit
$\lim_{\rho \to 0}f_n(\rho,\phi)$, which takes the form in \eqref{trigpol}. See also lemma
\ref{Slemma}.
Note that although $P_{a_1a_2\dots a_n}$ is analytic on $\hat V$, $f_n$ will not
be analytic in terms of $\xi,\eta$ since $\eta^a$ is direction dependent at $\Lambda \in S$.
In general we have the following lemmas:
\begin{lemma}\label{Slemma1}
Suppose that $f$ is an analytic function,
on a ball of radius $r_0$ around $\Lambda$ on $\hat V$. 
Then the restriction of $f$ to $S$, $f_L$,  can be decomposed as $f_L(\xi,\zeta)=f_1(\xi,\zeta)+i\rho
f_2(\xi,\zeta)$, where $f_1$ and $f_2$ are analytic in terms of $\xi$ and $\zeta$ 
on the disk $\xi^2+\zeta^2<\frac{r_0^2}{2}$, and where $\rho=\sqrt{\xi^2+\zeta^2}$.
Furthermore, if $f$ is real-valued then $f_1$ and $f_2$ are real-valued.
\end{lemma}
\begin{proof}
We start by splitting $f=f(x,y,z)$ into its even, $f_e$, and odd, $f_o$, part with respect to $z$.
We can now rewrite $f_e(x,y,z)=\tilde f_e(x,y,z^2)$ and $f_o(x,y,z)=z \tilde f_o(x,y,z^2)$,
where both $\tilde f_e$ and $\tilde f_o$ are analytic in their arguments (at least near 
$(0,0,0)$). The restriction of $\tilde f_e$ to $S$ gives $f_1$: $f_1(\xi,\zeta)=\tilde f_e(\xi,\zeta,-(\xi^2+\zeta^2))$,
while the restriction of $\tilde f_o$ gives $f_2$: $f_2(\xi,\zeta)=\tilde f_o(\xi,\zeta,-(\xi^2+\zeta^2))$. Adding these,
and also noting that $z \to i \rho$ gives the required decomposition.
On $S$ we have $|z|^2=\xi^2+\zeta^2$, hence $\xi^2+\zeta^2< \frac{r_0^2}{2}$ implies $x^2+y^2+z^2 < r_0^2$.
This gives the domain of analyticity. The reality follows from the construction.
\end{proof}

Remark: In $f_L$, the subscript $L$ stands for the 'leading term'.

Although tensor fields on $\hat V$ can be pulled back to $S \setminus \{\Lambda\}$, we will only need their
contractions with the appropriate number of $\eta^a$ vectors. This contraction will introduce a direction dependence
which shows up in the following lemma.

\begin{lemma}\label{Slemma}
Suppose $T_{a\ldots b}$ is an analytic tensor field on a ball of radius $r_0$ around $\Lambda$ on $\hat V$. 
Then the scalar field
$f_L=\eta^a \ldots \eta^b T_{a\ldots b}$ on $S$ can be written as $f_L(\xi,\zeta)=
\frac{1}{\rho^n}(f_1(\xi,\zeta)+i\rho f_2(\xi,\zeta))$, where $f_1$ and $f_2$ are analytic 
(on the disk with radius $\frac{r_0}{\sqrt{2}}$ around the origin) in terms of $\xi$ and $\zeta$,
and where $\rho=\sqrt{\xi^2+\zeta^2}$.
Furthermore, if $T_{a\ldots b}$ is real-valued then $f_1$ and $f_2$ are real-valued.
\end{lemma}
\begin{proof}
Consider the scalar field $g=x^a \ldots x^b T_{a\ldots b}$ on $\hat V$, where 
$$
x^a=x\left ( \frac{\partial}{\partial x}\right )^a+y\left ( \frac{\partial}{\partial y}\right )^a+z\left ( \frac{\partial}{\partial z}\right )^a .
$$
This scalar field is analytic on the same ball as $T_{a\ldots b}$. Hence we can use lemma \ref{Slemma1} 
to obtain analytic functions $f_1$ and $f_2$ such that $g_L=f_1+i\rho f_2$. The reality also follows from lemma \ref{Slemma1}.
Furthermore, $x^a=\rho\eta^a$ on $S$. Thus $g_L=\rho^n f_L$. But $f_L$ is bounded near the origin, 
thus we can divide $g_L$ by $\rho^n$ and get the lemma.
\end{proof}
We remark that the boundedness of  $f_L(\xi,\zeta)=
\frac{1}{\rho^n}(f_1(\xi,\zeta)+i\rho f_2(\xi,\zeta))$ when $\rho \to 0$ implies that
both $f_1$ and $f_2$ have zeros of sufficient order at $(\xi,\zeta)=(0,0)$. It also implies
that $f_L$ will be direction dependent there.

We can now contract \eqref{orgrec}
with $\eta^a$ and get the following theorem.
\begin{theorem}\label{scalrarrec2d}
Let $\hat V$ and $S$ be defined as in sections (\ref{tmoments}) and (\ref{momentsscalar}).
Let $\eta^a$ have the properties given by lemma \ref{etalemma}, and let $f_n$
be defined by \eqref{fn}.Then the recursion \eqref{orgrec} on $\hat V$ takes the form
\begin{equation}\label{scalarrec}
f_n=\eta^a\hat D_af_{n-1}
-\tfrac{(n-1)(2n-3)}{2}\eta^a\eta^b\widehat R_{ab}f_{n-2}
\end{equation}
on $S$. The moments of order $2^n$ are captured in the direction dependent limit 
$\lim_{\rho \to 0} f_n(\rho,\varphi)$.
\end{theorem}
\begin{proof}
That \eqref{orgrec} takes the form \eqref{scalarrec} 
 follows exactly as in \cite{backdahl2} using that $\eta^a\hat D_a \eta^b=0$, although the recursion is defined only on $S$ rather than on $\hat V$. The last statement is the content of
 lemma \ref{etalemma}c.
\end{proof}

\subsection{Simplified calculation of the moments}
In this section, we will show that it is possible to obtain the recursion \eqref{scalarrec}
without the term involving the Ricci tensor. This will be accomplished by using the conformal
freedom at hand, i.e. change $\Omega$. The conformal freedom is $\Omega \to
\tilde\Omega=\tfrac{\Omega}{\alpha}$ where $\alpha$ is analytic near $\Lambda$ with $\alpha(\Lambda)=1$\footnote{A more natural condition is the equivalent statement
$\Omega \to \tilde\Omega=\Omega \alpha$. However, this formulation gives slightly neater calculations.}.
$\hat D_a (\tfrac{1}{\alpha})$ at $\Lambda$ gives a shift of the moments which corresponds to a 
'translation' of the physical space, \cite{geroch}. Hence we can assume that $\hat D_a (\tfrac{1}{\gamma})=0$
at $\Lambda$.
It is to be noted that a change of $\Omega$ changes the (rescaled) potential
$\phi/\sqrt{\Omega}$. It also changes the normal coordinates on $\hat V$, and hence all conclusions
must be made with some care. In order to derive the simplified recursion \eqref{scalarreducedrec}, we will specify a $\alpha$ through $\alpha_L$, the restriction of $\alpha$ to $S$. However, in order to deduce that there exists a 
{\em real-valued} function $\alpha$ which the prescribed values of $\alpha_L$, we need to say more
on the representation of $\alpha_L$. This result and a useful estimate is the content of lemma \ref{serieslemma}.

\begin{lemma}\label{serieslemma}
Let $f_L=f_1(\xi,\zeta)+i\rho f_2(\xi,\zeta)$ where $f_1$ and $f_2$ are analytic 
on the ball $U=\{ |\xi|^2+|\zeta|^2<r_0^2 \}$, and where $\rho=\sqrt{\xi^2+\zeta^2}$.
We can then write
\begin{equation}\label{fcomplex}
f_L(\rho\cos\varphi,\rho\sin\varphi)=
\sum_{l=0}^\infty\sum_{m=-l}^l c_{l,m} e^{i m \varphi}\rho^l ,
\end{equation}
where 
\begin{equation}\label{complexconv}
\sum_{l=0}^\infty\sum_{m=-l}^l |c_{l,m}| \rho^l < \infty \; , \; \rho < r_0 .
\end{equation}
Furthermore, the converse is true; if $f_L$ is a function satisfying \eqref{fcomplex} and \eqref{complexconv}
then there are functions $f_1$ and $f_2$ analytic in $U$ such that $f_L=f_1+i\rho f_2$. 
The functions $f_1$ and $f_2$ are real-valued if and only if the coefficients
$c_{l,m}$ satisfy $\bar c_{l,m}=(-1)^{l-m}c_{l,-m}$.
\end{lemma}
\begin{proof}
Form the functions $f_1(\frac{x+y}2,\frac{x-y}{2i})$ and $f_2(\frac{x+y}2,\frac{x-y}{2i})$.\footnote{Although we will always take $\xi$ and $\zeta$ to be real, they are temporarily
complexified in this proof.}  These functions 
are analytic in terms of $x$ and $y$. Therefore there exist coefficients $c_{l,m}$ such that
\begin{equation}\label{f1f2}
f_j(\frac{x+y}2,\frac{x-y}{2i})=\sum_{l=0}^\infty\sum_{k=0}^l i^{1-j} c_{l+j-1,2k-l}x^ky^{l-k} ,
\end{equation}
where $j=1$ or $2$, and
$$
\sum_{l=0}^\infty\sum_{k=0}^l |c_{l+j-1,2k-l}||x|^k|y|^l < \infty \; , \; |\tfrac{x+y}2|^2+|\tfrac{x-y}{2i}|^2
=\tfrac{1}{2}(|x|^2+|y|^2) < r_0^2 .
$$
Now let $x=\rho e^{-i\varphi}$ and $y=\rho e^{i\varphi}$. For all $\rho^2=\tfrac{1}{2}(|x|^2+|y|^2) < r_0^2$ this shows that
\begin{equation}\label{flc}
f_L(\rho\cos\varphi,\rho\sin\varphi)=
\sum_{l=0}^\infty\sum_{k=0}^l c_{l,2k-l} e^{i (2k-l) \varphi}\rho^l +
\sum_{l=0}^\infty\sum_{k=0}^l c_{l+1,2k-l} e^{i (2k-l) \varphi}\rho^{l+1} .
\end{equation}
The reality condition follows easily from \eqref{f1f2} and \eqref{flc}.
A reorganisation gives the first part of the lemma. 
The converse is given by the substitution $x=\xi+i\zeta$ and $y=\xi-i\zeta$ in \eqref{f1f2}. The series 
converges absolutely for $|\xi|^2+|\zeta|^2=\rho^2 < r_0^2$.
\end{proof}
We now show that we can choose $\alpha$ such that the Ricci term in \eqref{scalarrec} vanishes.
\begin{lemma} \label{nnRnoll}
There exists a real-valued real analytic function $\alpha$ on $\hat V$ such that
the Ricci tensor $\tilde R_{ab}$ of the new metric $\tilde h_{ab}=\alpha^{-2}\hat h_{ab}$  
satisfies $\tilde \eta^a \tilde \eta^b \tilde R_{ab}=0$ on $\tilde S$, where both $\tilde \eta^a$, $\tilde S$
and the mapping $F$ are defined
in terms of coordinates which are normal with respect to the new metric $\tilde h_{ab}$.
\end{lemma}
\begin{proof} We first demonstrate that we can make $\eta^a \eta^b \tilde R_{ab}=0$ on $S$.
{}From \cite{wald}
$$
\tilde R_{ab}=\hat R_{ab}+\tfrac{1}{\alpha}\hat D_a \hat D_b \alpha+
\alpha^{-1}\hat h_{ab}\hat h^{cd}\hat D_c \hat D_d\alpha-
2\alpha^{-2}\hat h_{ab}\hat h^{cd}(\hat D_c \alpha) \hat D_d \alpha .
$$
On $S$, (using $\eta^a$ and $S$ belonging to $\hat h_{ab}$), this becomes
\begin{equation}\label{rkappa}
\eta^a \eta^b \tilde R_{ab}=
\eta^a\eta^b\hat R_{ab}+\tfrac{1}{\alpha_L}\eta^a\eta^b\hat D_a \hat D_b \alpha_L=
\eta^a\eta^b\hat R_{ab}+\tfrac{1}{\alpha_L}\tfrac{\partial^2 \alpha_L}{\partial\rho^2}
\end{equation}
where we have used $\eta^a \hat D_a \eta^b=0$. In order to make $\eta^a \eta^b \tilde R_{ab}=0$,
we require
\begin{equation} \label{omegaeq}
\alpha_{L \rho\rho}+\eta^a\eta^b\hat R_{ab}\alpha_L=0 .
\end{equation}
Note that, on $\hat V$, $\hat R_{ab}$ satisfies the conditions
of lemma \ref{Slemma}, therefore from lemma~\ref{Slemma} and lemma~\ref{serieslemma}
it follows that
\begin{equation}  \label{rserie}
\eta^a\eta^b\hat R_{ab}=\sum_{n=0}^\infty b_{n+2}(e^{i\varphi},e^{-i\varphi})\rho^n ,
\end{equation}
where each $b_n$ is a polynomial of degree at most $n$. 
It is easy to see that the reality condition from lemma \ref{serieslemma} is equivalent to
$\overline{b_{n}(e^{i\varphi},e^{-i\varphi})}=(-1)^n b_n(-e^{i\varphi},-e^{-i\varphi})$.
Hence the reality of $\hat R_{ab}$ implies this condition. Moreover, since
$\hat R_{ab}$ is analytic in a neighbourhood of $\Lambda$, there exists a $\rho_0$ such
that for any fixed $\varphi$, the series \eqref{rserie} converges for all $\rho < \rho_0$.
Well known ODE theory gives that for each fixed $\varphi$ there exists
a solution $\alpha_L$, analytic in $\rho$ for $0 \leq \rho<\rho_0$, i.e., for all $\varphi$ 
we have 
$$
\alpha_L=\sum_{n=0}^\infty a_n(\varphi)\rho^n , \quad \rho < \rho_0 .
$$
We note that $a_0=1$ while $a_1$
can be chosen to be 0 (translation). 
$\alpha_L$ will have the right regularity if we can show that each
$a_n(\varphi)$ is a polynomial in $e^{i\varphi}$ and $e^{-i\varphi}$ of
degree at most $n$. Equation \eqref{omegaeq} becomes
$$
-\sum_{n=0}^\infty n(n-1) a_n\rho^{n-2}=
\sum_{n=0}^\infty \sum_{j=0}^\infty b_{n+2} a_j\rho^{n+j}.
$$
Equating powers of $\rho$ we get
\begin{equation}\label{reckappa}
-(n+2)(n+1)a_{n+2}=\sum_{m=0}^n b_{m+2} a_{n-m},\quad n \geq 0 .
\end{equation}
The polynomials $a_0$ and $a_1$ has maximal degree $0$ respective $1$. The polynomials $b_n$ has maximal degree $n$.
Straightforward induction shows that $a_n$ has maximal degree $n$. 
Induction and \eqref{reckappa} also implies $\overline{a_n(e^{i\varphi},e^{-i\varphi})}=(-1)^n a_n(-e^{i\varphi},-e^{-i\varphi})$.
Thus, due to lemma \ref{serieslemma}, there are real-valued analytic functions $\alpha_1$ and $\alpha_2$ such that 
$\alpha_L=\alpha_1+i\rho \alpha_2$. The function $\alpha=\alpha_1(x,y)+z\alpha_2(x,y)$ is then a real-valued analytic 
extension of $\alpha_L$ to a ball in $\hat V$. This shows that there exist an $\alpha$
such that $\eta^a \eta^b \tilde R_{ab}=0$ on $S$. 

However, it then also follows that $\tilde \eta^a \tilde \eta^b \tilde R_{ab}=0$ on $\tilde S$.
{}From $\alpha$ we get $\tilde \Omega=\Omega/\alpha$, and the corresponding new metric $\tilde h_{ab}=\alpha^{-2}\hat h_{ab}$. 
Note that in \eqref{scalarrec}, where now $\eta^a\eta^b\widehat R_{ab}=0$,
the recursion is stated in terms of $\hat D_a$, i.e., it is expressed in terms of $\hat h_{ab}$
instead of $\tilde h_{ab}$. From $\tilde h_{ab}$
we get new normal coordinates, i.e., $(x,y,z)$ in $T_\Lambda \hat V_\mathbb C$,
are mapped into $\hat V_\mathbb C$ using the exponential map belonging to $\tilde h_{ab}$. We then construct $\tilde \eta^a$ and the mapping
$F$ with respect to these coordinates. Now, null geodesics, of which $S$ consists, are conformally invariant,
although they become non-affinely parametrised. This means that
$\tilde \eta^a \propto \eta^a$, and that points in $S$ are mapped into points in $\tilde S$.
Thus $\tilde \eta^a \tilde \eta^b\tilde R_{ab} \propto  \eta^a  \eta^b\tilde R_{ab}=0$ on $\tilde S$ (or $S$). 
\end{proof}

Henceforth we denote all entities defined via $\tilde h_{ab}$ instead of $\hat h_{ab}$ with
 a tilde. In particular, $\tilde D_a$ will denote the derivative operator associated with $\tilde h_{ab}$. Applying lemma \ref{nnRnoll} to theorem \ref{scalrarrec2d} we immediately get the following theorem.
\begin{theorem}\label{scalrarreducedrec2d}
Let $\hat V$ and $\tilde S$ be defined as in sections \eqref{tmoments} and \eqref{momentsscalar}, where
$\tilde S$ is defined in terms of normal coordinates connected to $\tilde h_{ab}$.
Let $\tilde \eta^a$ have the properties given by lemma \ref{etalemma} with respect to
$\tilde h_{ab}$, and let $\tilde f_n$
be defined by \eqref{fn} with $\tilde \eta^a$ replacing $\eta^a$.
Then the recursion \eqref{orgrec} on $\tilde V$ takes the form
\begin{equation}\label{scalarreducedrec}
\tilde f_n=\tilde \eta^a\tilde D_a\tilde f_{n-1}=(\tilde \eta^a \tilde D_a)^n \tilde f_0
=\frac{\partial^n}{\partial \tilde \rho^n}\tilde f_0
\end{equation}
on $\tilde S$. The moments of order $2^n$ are captured in the direction dependent limit 
$\lim_{\tilde \rho \to 0} \tilde f_n(\tilde \rho,\varphi)$.
\end{theorem}
\begin{proof}
Since $\tilde \eta^a \tilde D_a$ is $\frac{\partial}{\partial \tilde \rho}$, each $\tilde f_n$ is easily 
derived from $\tilde f_0$. 
Also, from lemma \ref{etalemma}c), we (again) know that the $2n+1$ degrees of freedom
of $P_{a_1 \ldots a_n}$ at $\Lambda$ are encoded in $\tilde f_n$.
\end{proof}

\section{Bounds on the moments} 
All multipole moments are encoded in $\tilde f_0$, and we note that the recursion
\eqref{scalarreducedrec} is identical to the recursion emanating from a scalar function
in $\mathbb{R}^3$ (after inversion). It is clear that many different functions on 
$\mathbb{R}^3$ will produce the same moments, but if we also require that the function, $g$ say,
is harmonic, $g$ is uniquely determined by the moments.

Thus, provided that we can connect a function which is harmonic in 
a neighbourhood of ${\mathbf 0} \in \mathbb{R}^3$
to each $\tilde f_0$, we have the following theorem: 

\begin{theorem}  \label{bounds}
Suppose that $(M,g_{ab})$ is a stationary asymptotically flat spacetime, admitting an analytic (rescaled) potential and an analytic
chart\footnote{As discussed before, the analyticity has been proved for the case with non-zero mass.}  on the conformally compactified manifold of timelike Killing 
trajectories, around the infinity point $\Lambda$.
Then there exist a (flat-)harmonic function $g$ in a neighbourhood of 
${\mathbf 0}\in T_{\Lambda}\hat V \cong \mathbb{R}^3$, such that all 
multipole moments of $M$ are given by
\begin{equation}\label{harmrec}
P_{a_1\dots a_n}(\Lambda)=(\nabla_{a_1} \dots \nabla _{a_n} g)(0) ,
\end{equation}
where $\nabla_a$ in the LHS of \eqref{harmrec} is the flat derivative operator in $\mathbb{R}^3$.
This puts a bound on the multipole moments, since the Taylor expansion 
$\sum_{|\alpha|\geq 0}\frac{\mathbf{r}^\alpha}{\alpha!}(\partial_\alpha g)(\mathbf 0)$
of  $g$ converges in a neighbourhood of  the origin in $\mathbb{R}^3$.
\end{theorem}

\begin{proof}
By lemma \ref{serieslemma} we have
$$
\tilde f_L(\rho\cos\varphi,\rho\sin\varphi)=
\sum_{l=0}^\infty\sum_{m=-l}^l c_{l,m} e^{i m \varphi}\rho^l, \quad \rho < r_0
$$
for some $r_0>0$, and we know that $\tilde f_L$ fully determines the moments of $M$.
We will now define the function $g$, which will be shown to be harmonic in a neighbourhood
of the origin. Finally we will argue that $g$ has the correct derivatives at $\mathbf 0$, i.e.,
that the equality \eqref{harmrec} is valid (for all $n \geq 0$).
First we define the coefficients 
$$
a_{l,m}=c_{l,m}i^{-m-l} 2^{1-l}\pi \frac{\sqrt{(l+m)!(l-m)!}}{\Gamma(l+\tfrac{1}2)\sqrt{2l+1}} .
$$
and the function
$$
g(r,\theta,\varphi)=\sum_{l=0}^\infty\sum_{m=-l}^l a_{l,m} Y^m_l(\theta,\varphi) r^l .
$$
Due to the construction, $g$ is harmonic at those (interior) points for which the
sum converges. We now establish convergence.
{}From the identity
$$
\sum_{m=-l}^l |Y_l^m(\theta,\varphi) |^2=\frac{2l+1}{4\pi}
$$
we get
\begin{gather*}
\left | \sum_{m=-l}^l a_{l,m}Y^m_l \right | \leq 
\sum_{m=-l}^l \left | a_{l,m}Y^m_l \right | \leq
\left ( \sum_{m=-l}^l | a_{l,m}|^2 \right )^{\frac{1}{2}}
\left ( \sum_{m=-l}^l |Y_l^m |^2 \right )^{\frac{1}{2}} \\
=\left ( \sum_{m=-l}^l | \sqrt{\tfrac{2l+1}{4\pi}} a_{l,m}|^2 \right )^{\frac{1}{2}}
=\frac{\sqrt{\pi}}{2^l\Gamma(l+\tfrac{1}2)} \left ( 
\sum_{m=-l}^l \left | c_{l,m} \sqrt{(l+m)!(l-m)!}\right |^2 \right )^{\frac{1}{2}}
\end{gather*}
Furthermore,
the inequality $(l+m)!(l-m!)\leq (2l)!$, when $-l\leq m\leq l$ follows from the convexity of $\ln(\Gamma(x))$.
Therefore
$$
\left | \sum_{m=-l}^l a_{l,m}Y^m_l \right | 
\leq \frac{\sqrt{\pi}\sqrt{(2l)!}}{2^l\Gamma(l+\tfrac{1}2)} 
\left ( \sum_{m=-l}^l | c_{l,m} |^2 \right )^{\frac{1}{2}}
$$
Next, the inequality $\frac{\pi (2l)!}{4^{l}\Gamma\left(l+\tfrac{1}2\right)^2}
=\frac{(2l)!!}{(2l-1)!!}=(2l+1)\frac{(2l)!!}{(2l+1)!!}\leq 2l+1$ gives
$$
\left | \sum_{m=-l}^l a_{l,m}Y^m_l \right | 
\leq \sqrt{2l+1}\left ( \sum_{m=-l}^l | c_{l,m} |^2 \right )^{\frac{1}{2}}
\leq \sqrt{2l+1}\sum_{m=-l}^l | c_{l,m} |
$$
But for all $\epsilon>0$ we have 
$\sqrt{2l+1}(1+\epsilon)^{-l} \rightarrow 0$ as  $l\rightarrow \infty$.
So the factor $\sqrt{2l+1}$ will not affect the radius of convergence.

Hence 
$$
 \sum_{l=0}^\infty\sum_{m=-l}^l a_{l,m}Y^m_l r^l \; \text{ converges if } r< r_0 .
$$
This shows that $g$ is well defined in a neighbourhood of $\mathbf 0$ in $T_\Lambda \hat V$,
and we must now show that we have equality in \eqref{harmrec}. We will do this by forming $g_L$
and then compare with $\tilde f_L$. Note, however, that $\tilde f_L$ is defined on $\tilde S \subset V_\mathbb{C}$,
while $g_L$ will be defined on the corresponding surface $\bar S=S_{\tilde h_{ab}(\Lambda)}$ in $T_\Lambda \hat V_\mathbb{C}$. (By $\bar S$ we denote the surface
defined by $F$ as previously, but where $F$ now maps $(\xi,\zeta)$ into $T_\Lambda \hat V_\mathbb{C}$ via the flat metric $\tilde h_{ab}(\Lambda)$.)
This means that $f_L$ and $g_L$
really can be compared only at $\Lambda$ (i.e. $\mathbf 0 \in T_\Lambda \hat V$), where also the equality \eqref{harmrec} is evaluated.
On the other hand, the radial derivatives of both entities are well defined and comparable at $\Lambda$.
In other words, if both $\tilde f_L$ and $g_L$ are equal when expressed in terms of the
coordinates $\xi, \zeta$, they will produce the same derivatives/moments. To summarise; in the case of $g$, 
the function $F$ and the vector $\eta^a$ are simply interpreted in $T_{\Lambda} \hat V_\mathbb{C}$
rather than in $V_\mathbb{C}$.

To proceed, we recall that 
$$
Y^m_l(\theta,\varphi)=i^{m-|m|}\sqrt{\frac{(2l+1)(l-|m|)!}{4\pi(l+|m|)!}}P^{|m|}_l(\cos\theta)e^{im\varphi}
$$
for $-l\leq m\leq l$, and that
$$
P^m_l(\cos\theta)=(-1)^m2^{-l} \sin^m\theta \sum_{k=0}^{\bigl \lfloor \tfrac{l-m}{2} \bigr \rfloor} 
\frac{(-1)^k (2l-2k)!}{k!(l-k)!(l-2k-m)!}\cos^{l-m-2k}\theta 
$$
where $0\leq \theta \leq \pi$ and $m\geq 0$.
Therefore,
$$
g(r,\theta,\varphi)=\sum_{l=0}^\infty\sum_{m=-l}^l \sum_{k=0}^{\bigl \lfloor \tfrac{l-|m|}{2} \bigr \rfloor}
\frac{ c_{l,m}\sqrt{\pi}(l-|m|)!(2l-2k)! e^{i m \varphi}\rho^{|m|} z^{l-|m|-2k} r^{2k}}
{i^{l-|m|}(-1)^k 4^lk!(l-k)!(l-|m|-2k)!\Gamma(l+\tfrac{1}2)} , 
$$
where $z=r\cos\theta$, $\rho=r\sin\theta$.
When we take the restriction of $g$ to $\bar S$, i.e., form $g_L$, only the terms with $k=0$ survives since $r_L=0$. Thus
$$
g_L(\rho\cos\theta,\rho\sin\theta)
=\sum_{l=0}^\infty\sum_{m=-l}^l
\frac{ c_{l,m}\sqrt{\pi}(2l)! e^{i m \varphi}\rho^l}
{4^l l!\Gamma(l+\tfrac{1}2)}
=\sum_{l=0}^\infty\sum_{m=-l}^l c_{l,m}e^{i m \varphi}\rho^l=\tilde f_L,
$$
which means that we have equality in \eqref{harmrec}.
\end{proof}

\section{Discussion}
In this paper we have studied the multipole moments of stationary asymptotically flat spacetimes. By using normal coordinates, and by exploiting the conformal freedom, we could show that the tensorial recursion \eqref{orgrec} could be replaced by the scalar
recursion \eqref{scalarreducedrec}. This recursion is a direction dependent
recursion on $\mathbb{R}^2$, where the moments are encoded in the direction dependent
limits at $\Lambda$.

Using this setup, we could also show that the multipole moments cannot grow too fast.
In essence, the rescaled potential behaves (locally) 
in the manner of a harmonic function on $\mathbb{R}^3$. The bounds on the moments given in theorem
\ref{bounds} gives the necessary part in a conjecture due to Geroch \cite{geroch},
and it is of course tempting to conjecture that this condition on the moments
also will be sufficient (as long as the monopole is real-valued).

Whether this can be proved using the techniques presented here is still an open question.

We also remark that similar questions concerning the convergence of asymptotic expansions in the static case are currently being studied by Friedrich, using a different technique, \cite{friedrich}.

\end{document}